\documentclass[preprint,prc,aps,showpacs,showkeys,groupedaddress,floatfix]{revtex4-1}
\usepackage{epsfig}
\usepackage{dcolumn}
\usepackage{bm}% bold mathhis
\usepackage[latin5]{inputenc}
\usepackage{graphics}
\usepackage{graphicx}
\usepackage{epsfig}
\usepackage{amssymb}
\usepackage{amsmath}

\begin{document}
\title{An Investigation of the Bound State Solutions of the Klein-Gordon Equation for the Generalized Woods-Saxon Potential in Spin Symmetry
and Pseudo-spin Symmetry Limits}
\author{B.C. L\"{u}tf\"{u}o\u{g}lu}
\affiliation{Department of Physics, Akdeniz University, 07058
Antalya, Turkey}
\date{\today}
\begin{abstract}
Recently, scattering of a Klein-Gordon  particle in the presence of mixed scalar-vector generalized symmetric Woods-Saxon potential was investigated for the spin symmetric and the pseudo-spin symmetric limits in one spatial dimension. In this manuscript, the bound state solutions of the Klein-Gordon equation with mixed scalar-vector generalized symmetric Woods-Saxon potential are examined analytically within the framework of spin and pseudo-spin symmetry limits. We prove that the occurrence of bound state energy spectrum exists only in the spin symmetric limit, while in the pseudo-spin symmetric limit, the bound state spectrum does not exist. Besides the theoretical proof, the Newton-Raphson numerical methods are used to calculate the bound state energy spectra of a neutral Kaon particle, confined in a generalized symmetric Woods-Saxon potential, energy well constituted with repulsive  or attractive surface interactions, for the spin and pseudo-spin symmetric limits, respectively. Numerical results are consistent with the  non-existence of the bound state energy spectrum in the pseudo-spin symmetric limit.
\end{abstract}
 \maketitle %%
\section{Introduction}\label{intro}

Symmetry is one of the basic and important concepts, which is often consulted to describe the laws of Nature. For instance, consequences of the spin symmetry (SS) and the pseudo-spin symmetry (PSS) have been an ongoing issue that is extensively discussed to characterize the nuclear structure phenomena in  nuclear physics e.g. magic numbers \cite{RefHaxel1949}, deformation and superdeformation \cite{RefBohr1982, RefDudek1987}, identical rotational bands \cite{RefByrskietal1990}, magnetic moments \cite{RefNazarewiczetal1990}. Initially, these symmetries are asserted by Smith et al. \cite{RefSmithTassie1971}  and Bell et al. \cite{RefBellRuegg1975} in their independent studies. Afterwards, the subject became more and more  popular and was studied by many authors \cite{RefMottelson1991, RefGinocchio1999, RefSugawaraetal2000, RefZhouMengRing2003, RefMishustinetal2005, RefCastro2005, RefAlhaidariBahlouliAl-Hasan2006, RefGuo2006, RefAlbertoCastroMalheiro2007, RefLuetal2012, RefLianZhaogetal2011, RefXu2012, RefHamzaviRijabi2013,  RefAlbertoCastroMalheiro2013, RefLiangShen2013, RefAlbertoMalheiroFredericoCastro2015, RefIkotetal2015, RefAlbertoetal2016, RefLuLiuetal2017}.  Detailed reviews of using these symmetries in nuclear structure phenomena, including the open problems, were given by Ginocchio \cite{RefGinocchio2005}  and recently  by Liang et al.  \cite{RefLiangMengZhou2015}. Furthermore, the foundations of these symmetries were  investigated comprehensively \cite{RefBahrietal1992, RefBlokhinBahriDraayer1995, RefGinocchio1997}. Blokhin et al. employed a helicity transformation on a non-relativistic single particle Hamiltonian and proved that the reduced radial wave function's asymptotic behaviours differ from the former cases, resulting in the increase  the diffuseness  \cite{RefBlokhinBahriDraayer1995}. Only a few years before the work of Blokhin,  Bahri et al. performed a similar transformation on non-relativistic harmonic oscillator \cite{RefBahrietal1992}. They concluded the existence of certain conditions that require having a PSS  in the Hamiltonian,  which is consistent with relativistic mean field approaches. In order to explain this correlation,
Ginocchio claimed that PSS is a symmetry that occurs in the presence of an attractive scalar potential, $V_s$, and a repulsive vector potential, $V_v$ together  with nearly equal magnitudes, $V_v+V_s=\varepsilon^-$, in relativistic mean field theory \cite{RefGinocchio1997}. Meng et al. proved that the PSS becomes an exact symmetry when $\frac{d \varepsilon^-}{dr}=0$ \cite{RefMengetal2013}. Moreover, they imitated the PSS as a competition between  the pseudo-centrifugal barrier and the pseudo-orbital potential in  real nuclei. Contrary to the PSS, the SS is defined by  $V_v-V_s=\varepsilon^+$.  In finite nuclei, the constants $\varepsilon^+$ and $\varepsilon^-$ are zero \cite{RefGinocchio1999}. Note that, Ginocchio attributed the mechanism behind these symmetries to the Dirac equation. Alberto et al. discussed the required conditions to obtain
equivalent energy spectra of relativistic $spin-\frac{1}{2}$ and $spin-0$ particles in the existence of mixed vector and scalar potential energies \cite{RefAlbertoCastroMalheiro2007}. They concluded that those conditions do not depend on the potential parameters, thus, the shape of the potential energies, but  just on  whether their difference (SS) or their sum (PSS) differ by a constant.

The solutions of the Dirac and the Klein-Gordon(KG) equations are obtained by using these symmetries in the presence of various potential energies,
e.g. SS \cite{RefGinocchio2004} and PSS  \cite{RefLisboaetal2004, RefGuoetal2005} in the relativistic harmonic oscillator potential,
resonant states solutions and PSS in the Dirac-Morse potential energy \cite{RefLiuetal2013}, PSS in the Dirac equation with a Lorentz structured Woods-Saxon potential \cite{RefAlbertoetal2001}, SS and PSS in the Hulth$\acute{e}$n-like potential and tensor interaction \cite{RefHamzaviIkhfdair2013},  SS scattering state solutions of KG particles by $q-$Parameter Hyperbolic P\"oschl-Teller potential \cite{RefIkotAtAll2015}, SS and PSS scattering of KG particles with generalized symmetric Woods-Saxon potential \cite{RefLutfuogluJiriJan2018}.

Among those studies, the Woods-Saxon potential (WSP)  is a well-known potential energy which has been used to describe the differential cross-section of the protons elastic scattering from medium and heavy nuclei in the optical model, replacing the  square well potential \cite{RefWoodsSaxon1954}.
WSP energy basically depends on three parameters: the depth of the potential well, the diffusion and the effective well radius. In the literature, solutions of the WSP in the  non-relativistic \cite{BookFlugge} and the relativistic \cite{RefKennedy2002, RefGuoetal2002, RefRojasVillalba2005} cases exist. Besides these studies, the modification of the WSP energy is examined in many  articles where  the potential is deformed with two additional parameters \cite{RefAlpdoganetal2013, RefAlpdoganhavare2014}. Such modifications are sometimes named as the generalization of the WSP by their authors. This, however,  should not be perceived as an exact generalization, since the characteristic shape of the potential does not change and the so-called generalized potential energy does not include additional physical effects. Another, "generalization" is defined with an additional physical term corresponding to  surface interactions. Satchler proposed this extra term, which is linearly proportional to the spatial derivative of the WSP, corresponding to the forces that a nucleon suffers attractively or repulsively in the vicinity of the effective radius \cite{BookSatchler}. This "Generalized Woods-Saxon potential" has been investigated comprehensively \cite{RefKobosetal1982, RefKouraYamada2000, RefBoztosun2002, RefBerkdemirBerkdemir2005, RefBoztosunetal2005, RefGonulKoksal2007, RefTianWangLi2007, RefBadalovAhmado2009,  RefKocaketal2010, RefDapoetal2012, RefIkotAkpan2012, RefIkhdairFalayeHamzavi2013, RefCandemirBayrak2014, RefBayrakSahin2015, RefBayrakAciksoz2015, RefCapakPetrellis2015, RefLiendoCastro2016, RefCapakGonul2016}.

 In this manuscript, one-dimensional form of this generalized potential that possesses parity invariance is studied. Thus, from now on it will be called as generalized symmetric Woods-Saxon potential (GSWSP) energy. Recently scattering, tight and quasi-bound solutions of the GSWSP in Schr\"odinger equation are given by \cite{RefLutfuogluetal2016}. Moreover, the effects of the surface interactions to the thermodynamics of a nucleon are examined via the non-relativistic \cite{RefLutfuoglu2018} and the relativistic \cite{RefLutfuogluCJP2018} approaches. Lately, scattering of a KG particle in the GSWSP in the presence of SS and PSS  is investigated \cite{RefLutfuogluJiriJan2018}. The motivation of this manuscript is to examine the bound state solutions of the KG equation for the GSWSP energy in the limits of the SS and PSS, respectively. Surprisingly, the constraints on the wave numbers allow us to obtain the energy spectrum of the bound state in the SS limit, while they do not allow bound states in the PSS limit.

The layout of the manuscript is as follows.  In
section~\ref{sec:gsws} the GSWSP is introduced, where, the
dependence of the potential on the parameters and the surface
interactions are qualitatively discussed by plotting them with
arbitrarily chosen parameters. In section~\ref{sec:kg} the
time-independent KG equation in the SS and PSS limits are derived
briefly. In section~\ref{sec:baglidurumcozumu}, the most general
solution for the bound states is derived in the SS and PSS limits.
Four subsections are given
 in this section. In subsection~\ref{Bound State Conditions}, the constraints on the wave numbers are investigated comprehensively and the first
 remarkable result of the manuscript is given by proving the non existence of the PSS solution. In subsection~\ref{ASB}, the asymptotic behaviors
 are used to eliminate the non-physical solutions from the general solutions. In following subsections, \ref{CC} and \ref{Quan}, the continuity
 conditions are employed to determine the quantized energy spectra and their corresponding wave functions in the SS limit.
 In section~\ref{sec:uygulama}, as a  numerical application, a neutral Kaon is considered and its energy spectra are calculated for potentials
 with repulsive and attractive surface terms, by adjusting parameters. The corresponding wave functions, the probability densities and the
 dependence of the energy spectra on the SS limit parameter are visually revealed.  Moreover, the theoretical proof of the non-existence
 of the PSS limit is verified numerically. Finally, the conclusions are given in section~\ref{sec:sonuc}.

\section{Generalized Symmetric Woods-Saxon Potential}\label{sec:gsws}

The GSWSP that is under investigation in this manuscript has the following form \cite{RefLutfuogluJiriJan2018}

\begin{eqnarray}\label{gws}
  V_{GSWSP}(x)&=&\theta{(-x)}\Bigg[-\frac{V_0}{1+e^{-\alpha(x+L)}}+ \frac{We^{-\alpha(x+L)}}{\big(1+e^{-\alpha(x+L)}\big)^2}\Bigg]\nonumber  \\
  &&+ \theta{(x)}\Bigg[-\frac{V_0}{1+e^{\alpha(x-L)}}+  \frac{We^{\alpha(x-L)}}{\big(1+e^{\alpha(x-L)}\big)^2}\Bigg]. \label{GSWSP}
 \end{eqnarray}
Here $\theta{(\pm x)}$ are the Heaviside step functions. The GSWSP depends on four parameters. Three of these parameters are common with the WSP: $V_0$ determines the depth of the well,  $\alpha$ is the reciprocal diffusion constant and  $L$ is the effective radius. Note that these three parameters are positive and real numbers in this manuscript. The fourth parameter, $W$ , is the measure of the surface interactions and can be negative or positive depending on the physical problem, i.e.  a repulsive or an attractive surface effect is adjusted with the proportionality constant to be negative or positive. Note that  in the presence of the surface effects the potential energy
changes, but a \emph{"pocket"} for the repulsive  and a \emph{"barrier"} for the attractive surface effects appears  if and only if $|W| > V_0$ with
the depth or height equals to  $\frac{(V_0-W)^2}{4W}$. Moreover, $W$ is linearly proportional to the former three parameters, and its value  can be
 determined by  the conservation laws such as momentum and energy conservations in a real problem.

\begin{figure}[!htb]
\centering
\includegraphics[totalheight=0.45\textheight]{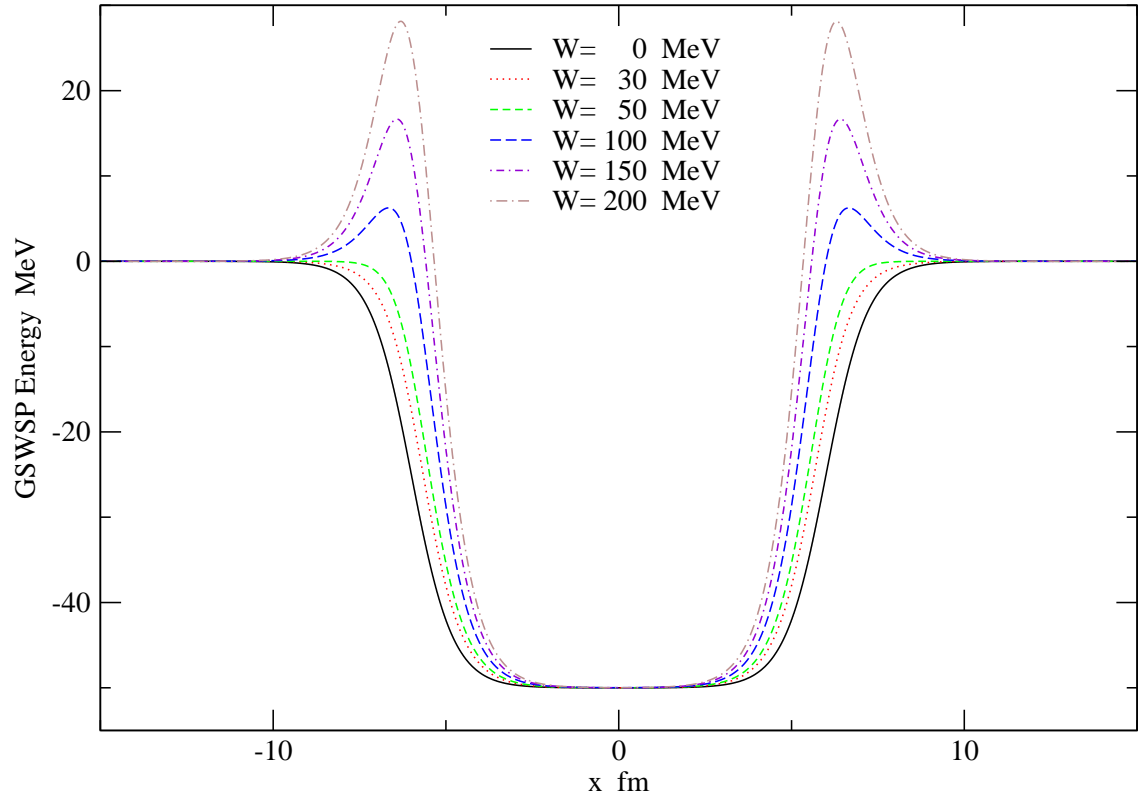}
   \caption{GSWSP energy wells with arbitrary determined potential parameters $V_0=50$ $MeV$, $\alpha=1.\bar{6}$ $fm^{-1}$ and $L=6$ $fm$. Repulsive
   surface forces become dominant with the increase of the $W$ parameter.} \label{fig1:2016_10_08_can_pot_v_01_pozitif_yuzey_terimi_icin}
\end{figure}

\begin{figure}[!htb]
\centering
\includegraphics[totalheight=0.45\textheight]{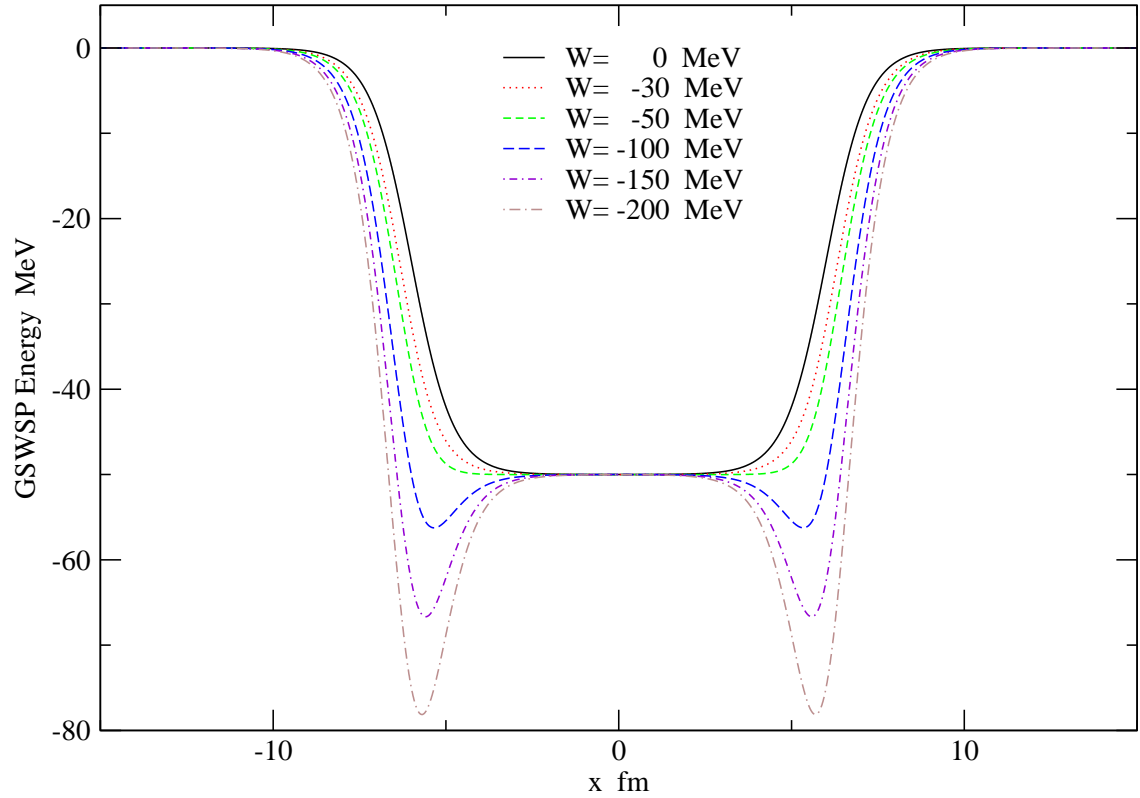}
   \caption{GSWSP energy wells with arbitrary determined potential parameters $V_0=50$ $MeV$, $\alpha=1.\bar{6}$ $fm^{-1}$ and $L=6$ $fm$.
   Attractive surface forces become dominant with the negative increase of the $W$ parameter.}
   \label{fig2:2016_10_08_can_pot_v_01_negatif_yuzey_terimi_icin}
\end{figure}

\begin{figure}[!htb]
\centering
\includegraphics[totalheight=0.45\textheight]{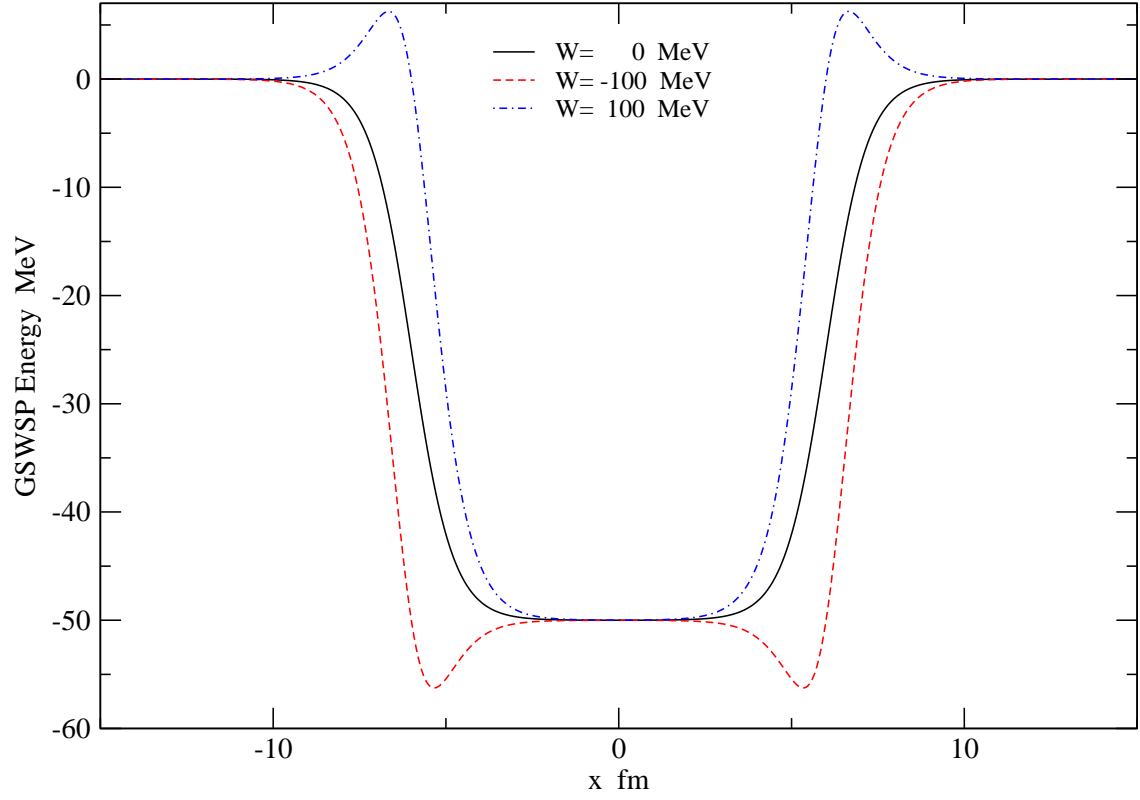}
   \caption{The parameters $V_0=50$ $MeV$, $\alpha=1.\bar{6}$ $fm^{-1}$ and $L=6$ $fm$ are used to plot potential energy wells. The solid line
   represents the WSP energy, where $W=0$, indicates no surface effects are taken into account. The dashed dot line (blue) and the dashed line (red)
   indicate  repulsive and attractive surface forces are considered in GSWSP energy well, respectively.  Note that, a pocket in the attractive case
   and a barrier in the repulsive case occurs, since $V_0<|W|$. The pocket depth and the barrier height are calculated to be $-56.25$ $MeV$ and
   $6.25$ $MeV$, respectively.} \label{fig3:2016_10_08_can_pot_gercekte_ne_oluyor}
\end{figure}

\begin{figure}[!htb]
\centering
\includegraphics[totalheight=0.45\textheight]{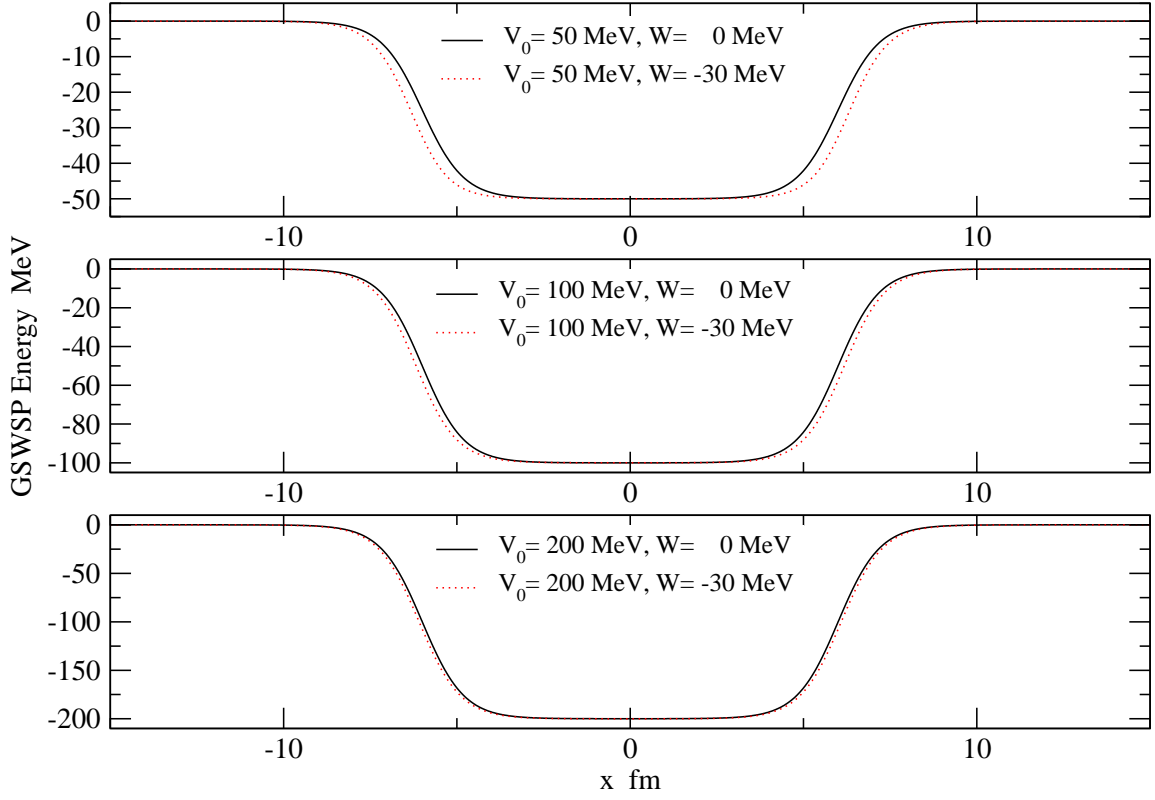}
   \caption{GSWSP well shade off into WSP well when the potential depth parameter $V_0$ is much greater then the absolute value of the surface
   effects coefficient $W$. The constant parameters are chosen to be  $\alpha=1.\bar{6}$ $fm^{-1}$ and  $L=6$ $fm$.}
   \label{fig4:2016_10_08_can_pot_v_02_negatif_yuzey_terimi_icin}
\end{figure}

\begin{figure}[!htb]
\centering
\includegraphics[totalheight=0.45\textheight]{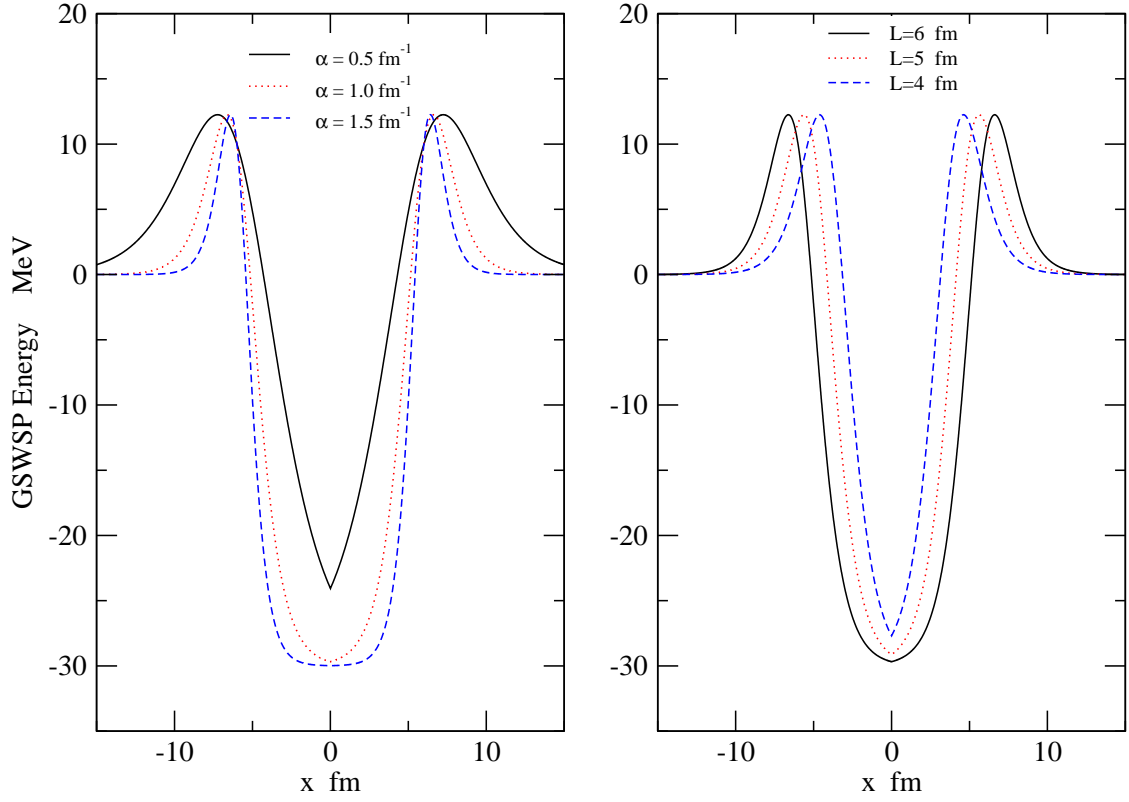}
   \caption{ Variation of the GSWSP energy versus  the reciprocal diffusion parameter and the nuclear radius with the repulsive surface effects.
   Here, the potential parameters are chosen to be $V_0=30$ $MeV$, $W=100$ $MeV$ for both graphs and $L=6$ $fm$ for the left-hand side while
   $\alpha=1$  $fm^{-1}$ for right-hand side.} \label{fig5:Potential_GSWS_a_and_L_varies}
\end{figure}

To have an exact realization of the solutions of the bound state of the KG equation for the GSWSP energy, it is necessary to investigate the
 potential energy qualitatively, especially the surface interaction term dependence. Therefore, the generalization of the usual WSP energy with the increasing repulsive and attractive surface forces are plotted versus spatial distance in Fig.~\ref{fig1:2016_10_08_can_pot_v_01_pozitif_yuzey_terimi_icin}  and  in Fig.~\ref{fig2:2016_10_08_can_pot_v_01_negatif_yuzey_terimi_icin}, respectively. One sees, then, how the potential barrier or pocket occurs as the consequence of the presence of surface interactions only after $|W|$ exceeds the $V_0$ value as predicted. Besides this, in the repulsive case, an increase of $W$ ends up in the squeeze of the potential well, while in the attractive case, a decrease of $W$ causes the well enlargement. These effects are shown in comparison to the WSP well in Fig.~\ref{fig3:2016_10_08_can_pot_gercekte_ne_oluyor}. Hence, an upward shift in the energy spectra is expected for the repulsive interaction, while a downward shift is expected in the attractive case.  On the other hand, an increase of the rate of the bulk effect to the surface effect is investigated in Fig.~\ref{fig4:2016_10_08_can_pot_v_02_negatif_yuzey_terimi_icin}.  One finds that,   when the bulk effect becomes more dominant than  the surface effect,  the GSWSP tends to resemble the usual WSP. Finally, the dependence of the other potential parameters on the GSWSP is investigated in Fig.~\ref{fig5:Potential_GSWS_a_and_L_varies}. It is observed that the GSWSP cannot be a smooth potential energy in some critical values of the reciprocal diffusion parameter and the effective well radius. Therefore, appropriate parameters satisfying $e^{\alpha L} >>1$ condition are chosen  in Section~\ref{sec:uygulama} to calculate the energy spectrum.

\section{The Klein-Gordon equation} \label{sec:kg}
The KG equation is a linear homogeneous second order partial equation
\begin{eqnarray}
  \Big[\hat{p}^\mu \hat{p}_\mu- (m_0 c)^2\Big]\Phi(\vec{r},t) &=& 0
\end{eqnarray}
and describes spinless scalar or pseudo-scalar particle dynamics. It is relativistically invariant and  was proposed by  Klein \cite{Klein1926} via
two quantities,  rest mass and linear momentum. Here, the speed of light is denoted by $c$ while  the four-momentum operator is represented with $\hat{p}^\mu$.

The interactions of a KG particle with the electromagnetic field are described by the minimal substitution with an electromagnetic coupling term $e$
where $e$ is a real number.
\begin{eqnarray}
  \hat{p}^\mu &\rightarrow & \hat{p}^\mu - \frac{e}{c} A^\mu
\end{eqnarray}
The four-vector  potential $A^\mu$ contains time and spatial components. In this manuscript the spatial potential terms are chosen to be zero. The
non-zero time component term is called "a vector potential", $e A^0\equiv V_v$. On the other hand, "a scalar potential", $V_s$, is coupled to the rest
 mass with another coupling constant $g$.
\begin{eqnarray}
  m_0 &\rightarrow & m_0 + \frac{g}{c^2}V_s.
\end{eqnarray}
In this paper, a time independent potential energy is going to be used  in  $(1+1)$ Minkowski space-time. In the weak regime, where $g<<1$, the mass
coupling drops and the expression is reduced to the usual KG equation. In the strong regime, where $g\approx 1$, the time independent KG equation
becomes
\begin{eqnarray}
  \Bigg[\frac{d^2 }{d x^2}+\frac{1}{\hbar^2c^2}\Big[ \big(E -V_v \big)^2- \big(m_0c^2 + g V_s\big)^2\Big]\Bigg]\phi(x) .
  &=& 0, \,\,\,\,\,\,\,\,\,\,\,\,\,\,\,\,\,\,\,\, \label{KGham}
\end{eqnarray}
Here $\hbar$ is the Planck constant. The relations between the potential energies are given in the SS limit as
\begin{eqnarray}
  V_v -  g V_s&=& \varepsilon^+.
\end{eqnarray}
Then Eq.~(\ref{KGham}) yields to
\begin{eqnarray}
   \Bigg[\frac{d^2 }{d x^2}+\frac{1}{\hbar^2c^2} \bigg[E^2-\big(\varepsilon^+ -m_0c^2\big)^2 -2V_v\Big(E-\big(\varepsilon^+ -m_0c^2\big)\Big)\bigg]
   \Bigg]\phi(x) &=& 0.  \,\,\,\,\,\,\,\,\,\,\,\,\,\,\,\,
\end{eqnarray}
Similarly, for the PSS limit
\begin{eqnarray}
  V_v + g V_s&=& \varepsilon^-,
\end{eqnarray}
gives
\begin{eqnarray}
  \Bigg[\frac{d^2 }{d x^2}+\frac{1}{\hbar^2c^2} \bigg[E^2-\big(\varepsilon^- +m_0c^2\big)^2 -2V_v\Big(E-\big(\varepsilon^- +m_0c^2\big)\Big)\bigg]
  \Bigg]\phi(x) &=& 0.  \,\,\,\,\,\,\,\,\,\,\,\,\,\,\,\,
\end{eqnarray}
It is worth noting that for  finite nuclei, the constants $\varepsilon^+$ and $\varepsilon^-$ are shown to  be  zero~\cite{RefGinocchio1999}.
In this paper, the KG equation that possesses SS or PSS is examined by the common expression  given by
\begin{eqnarray}
    \Bigg[\frac{d^2 }{d x^2}+\frac{1}{\hbar^2c^2}\Big[ \Big(E^2-\big(\varepsilon^{\mp}\pm m_0c^2\big)^2\Big)-2V_v\Big(E- \big(\varepsilon^{\mp}
    \pm m_0c^2\big)\Big)\Big]\Bigg]\phi^{\mp}(x) &=& 0.
  \,\,\,\,\,\,\,\,\,\,\,\,\,\,\,\,\,\,\,\,\label{KG1}
\end{eqnarray}
Here $\varepsilon^{\mp}$ is used, $+$ indicates the SS, while $-$  represents the PSS limits.

\section{Bound State Solutions}\label{sec:baglidurumcozumu}

The GSWSP energy has parity invariance in one dimension. The solutions in either direction can be deduced by exploiting this symmetry. Here, a solution for negative $x$ is analyzed, and then, the extension of the solution in the positive $x$ region will be given.

The substitution of the GSWSP given in Eq.~(\ref{GSWSP})  to Eq.~(\ref{KG1}) gives,
  \begin{eqnarray}
  \Bigg[\frac{d^2}{dx^2}+\alpha^2\bigg[-{\epsilon_\mp}^2+\frac{{\beta_\mp}^2}{1+e^{-\alpha(x+L)}}+\frac{{\gamma_\mp}^2}
  {\big(1+e^{-\alpha(x+L)}\big)^2}\bigg] \Bigg]\phi_L^{\mp}(x) &=& 0. \label{KG2x<0}
\end{eqnarray}
Here, the following parameter abbreviations are used to reveal the most general dependence of the parameters.
\begin{eqnarray*}
-{\epsilon_\mp}^2 &\equiv& \frac{E^2-\big(\varepsilon^{\mp}\pm m_0c^2\big)^2}{\alpha^2\hbar^2c^2},\\
{\beta_\mp}^2  &\equiv& \frac{2\Big(E- \big(\varepsilon^{\mp}\pm m_0c^2\big)\Big)(V_0-W)}{\alpha^2\hbar^2c^2},  \\
{\gamma_\mp}^2    &\equiv& \frac{2\Big(E- \big(\varepsilon^{\mp}\pm m_0c^2\big)\Big)W}{\alpha^2\hbar^2c^2}.
\end{eqnarray*}
Remark that $\varepsilon^{\mp}$ is zero for the GSWSP energy. A new transformation on the spatial component is defined as
\begin{eqnarray}
  z &\equiv& \Big[1+e^{-\alpha(x+L)}\Big]^{-1},
\end{eqnarray}
and it is found out that the KG equation yields
\begin{eqnarray}
&&\Bigg[\frac{d^2}{dz^2}+\bigg(\frac{1}{z}+\frac{1}{z-1}\bigg)\frac{d}{dz}+\bigg(\frac{\beta_\mp^2-2\epsilon_\mp^2}{z}
-\frac{\epsilon_\mp^2}{z^2}-\frac{\beta_\mp^2-2\epsilon_\mp^2}{(z-1)}+\frac{\beta_\mp^2+\gamma_\mp^2-\epsilon_\mp^2}
{(z-1)^2}\bigg)\Bigg]  \phi_{L}^{\mp}(z) = 0. \,\,\,\,\,\,\, \label{KG3x<0}
\end{eqnarray}
Note that, the boundaries are mapped into $z\rightarrow 0 (x\rightarrow -\infty)$ and $z\rightarrow 1 (x\rightarrow 0)$ since $e^{\alpha L} >> 1$. The general solution is considered with an ansatz as
\begin{eqnarray}
  \phi_{L}^{\mp}(z) &\equiv& z^{\mu_\mp} (z-1)^{\nu_\mp}f_\mp(z), \label{gensol}
\end{eqnarray}
where $\mu_\mp$ and $\nu_\mp$ satisfy
\begin{eqnarray}
  \mu_\mp^2-\epsilon_\mp^2 &=& 0, \\
  \nu_\mp^2+ \beta_\mp^2+\gamma_\mp^2-\epsilon_\mp^2 &=& 0,
\end{eqnarray}
conditions. Note that  positively defined wave numbers $k_\mp$
\begin{eqnarray}
k_\mp&\equiv&\frac{1}{\hbar c}\sqrt{-\big(E\mp m_0c^2\big)\big(E\pm m_0c^2 \big)}, \label{k1}
\end{eqnarray}
and $\kappa_\mp$
\begin{eqnarray}
\kappa_\mp &\equiv& \frac{1}{\hbar c} \sqrt{\big(E\mp m_0c^2\big)\big(E\pm m_0c^2+2V_0 \big)}\, , \label{k2}
\end{eqnarray}
are related with the following parameters
\begin{eqnarray}
  \mu_\mp  &=& \frac{k_\mp}{\alpha}, \\
  \nu_\mp&=& \frac{i\kappa_\mp}{\alpha}.
\end{eqnarray}
Substituting Eq.~(\ref{gensol}) into the Eq.~(\ref{KG3x<0}) gives
\begin{eqnarray}
   &&z(1-z)f_\mp{''}(z)+ \Big[(1+2\mu_\mp)-(1+2\mu_\mp+2\nu_\mp+1)z\Big]f_\mp{'}(z)\nonumber \\&&-\Big[(\mu_\mp+ \nu_\mp)^2+(\mu_\mp+\nu_\mp)+
   \gamma_\mp^2\Big]f_\mp(z)= 0. \label{KG_eq5}
\end{eqnarray}
Notice that the resulting differential equation has the very well known form of the Hypergeometric equation
\begin{eqnarray}
   z(1-z)u{''}(z)+ \Big[c-(1+a+b)z\Big]u'(z)- ab u(z) &=& 0, \label{Hypergeometric_dif_eq}
\end{eqnarray}
that has  solutions
\begin{eqnarray}
  u(z) &=& A\,\,\, {}_2F_1[a,b,c;z]+B z^{1-c}\,\,\, {}_2F_1[1+a-c,1+b-c,2-c;z]\,, \label{hipergeometrik_genel_cozum}
\end{eqnarray}
where ${}_2F_1$ is a hypergeometric function. Comparing Eq.~(\ref{KG_eq5}) with Eq.~(\ref{Hypergeometric_dif_eq}) the solution of the function is
determined as follows
\begin{eqnarray}
  f_\mp(z) &=& D_1^\mp   \,\,\, {}_2F_1[\mu_\mp+\theta_\mp+\nu_\mp,1+\mu_\mp-\theta_\mp+\nu_\mp,1+2\mu_\mp;z]\nonumber \\
    &+& D_2^\mp z^{-2\mu_\mp}  \,\,\, {}_2F_1[-\mu_\mp+\theta_\mp+\nu_\mp,1-\mu_\mp-\theta_\mp+\nu_\mp,1-2\mu_\mp;z] \,, \,\,\,\,\,\,\,\,
\end{eqnarray}
where
\begin{eqnarray}
\theta_\mp \equiv \frac{1}{2}\mp \sqrt[]{\frac{1}{4}-\gamma_\mp^2}.
\end{eqnarray}
Hence, the most general solution in the negative region is found to be
\begin{eqnarray}
  \phi_L^\mp(z) &=& D_1^\mp z^{\mu_\mp} (z-1)^{\nu_\mp} \,\,\, {}_2F_1[\mu_\mp+\theta_\mp+\nu_\mp,1+\mu_\mp-\theta_\mp+\nu_\mp,1+2\mu_\mp;z]
  \nonumber \\
  &+&D_2^\mp z^{-\mu_\mp} (z-1)^{\nu_\mp} \,\,\, {}_2F_1[-\mu_\mp+\theta_\mp+\nu_\mp,1-\mu_\mp-\theta_\mp+\nu_\mp,1-2\mu_\mp;z]. \,\,\,\,\,\,\,\,
\end{eqnarray}
The covariance of the KG equation implies that $\phi_R^\mp(y)$ will be symmetric to $\phi_L^\mp(z)$.
\begin{eqnarray}
  \phi_R^\mp(y) &=& D_3^\mp y^{\mu_\mp} (y-1)^{\nu_\mp} \,\,\, {}_2F_1[\mu_\mp+\theta_\mp+\nu_\mp,1+\mu_\mp-\theta_\mp+\nu_\mp,1+2\mu_\mp;y]\nonumber
  \\
  &+&D_4^\mp y^{-\mu_\mp} (y-1)^{\nu_\mp} \,\,\, {}_2F_1[-\mu_\mp+\theta_\mp+\nu_\mp,1-\mu_\mp-\theta_\mp+\nu_\mp,1-2\mu_\mp;y]. \,\,\,\,\,\,\,\,
\end{eqnarray}
Here $D_1^\mp, \cdots, D_4^\mp $ represents four normalization constants  and
\begin{eqnarray}
  y \equiv \Big[1+e^{\alpha(x-L)}\Big]^{-1},
\end{eqnarray}
is used  for the coordinate transformation in positive region. Analogously to the case in negative region,  the boundaries in the positive region are also mapped into $y\rightarrow 0 (x\rightarrow \infty)$ and $y\rightarrow 1 (x\rightarrow 0)$ since $e^{\alpha L} >> 1$.

\subsection{Bound State Conditions}\label{Bound State Conditions}
Since the KG particles are confined, their wave functions should exponentially decay outside  the potential well, whereas, sinusoidal wave functions should accompany to the particles within the well. We investigate these conditions comprehensively for the SS and PSS limits. Two wave numbers defined in  Eq.~(\ref{k1}) and  Eq.~(\ref{k2}) should be real to satisfy the confinement conditions
\begin{eqnarray}
   -\big(E\mp m_0c^2\big)\big(E\pm m_0c^2 \big) &>&0, \label{SSbirincikosul}\\
  \big(E\mp m_0c^2\big)\big(E\pm m_0c^2+2V_0 \big)&>&0, \label{SSikincikosul}
\end{eqnarray}
in addition to the condition $V_0>0$. Note that, due to Klein paradox, $V_0$ has an upper limit.  This value is in order of $mc^2$ of the depth parameter and the bound particle's energy reaches the continuum. This is known as supercriticality \cite{Calogeracos1999}.

\subsubsection{Spin symmetric limit}

\begin{figure}[!htb]
\centering
\includegraphics[totalheight=0.45\textheight]{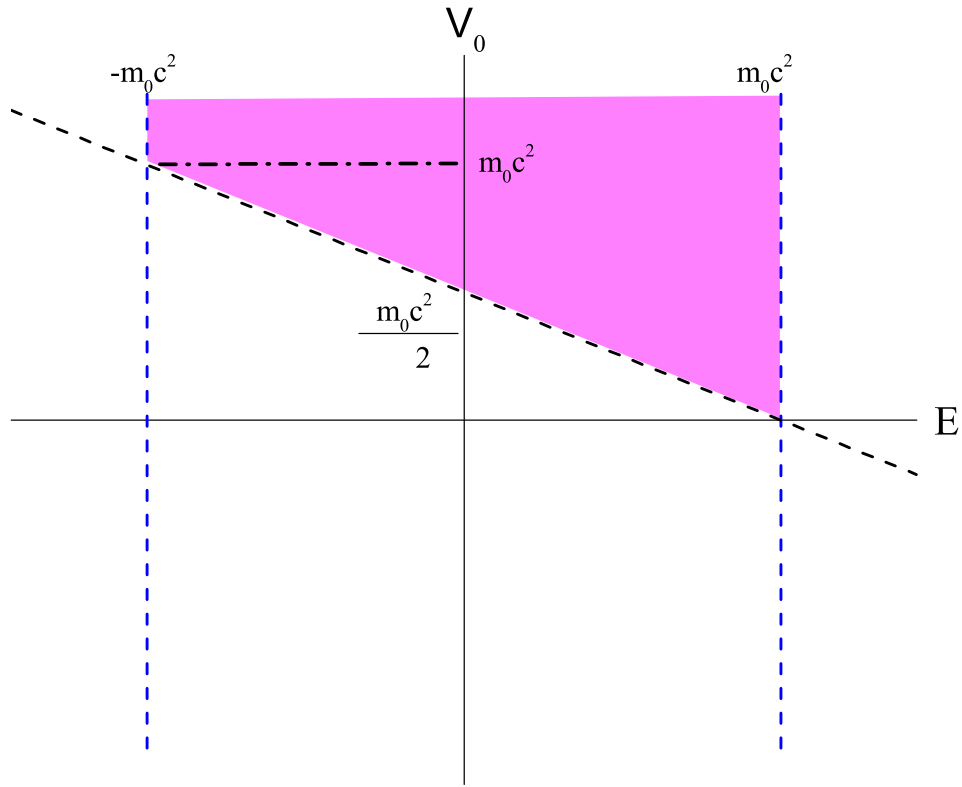}
   \caption{Possible energy eigenvalue region for a confined particle in  SS limit. Potential well depth parameter and rest mass energy determine to  have positive and/or negative eigenvalues.} \label{fig6:SSpotandenergyrelations}
\end{figure}

In order to comprehend the inequalities, that are found as the bound state conditions, we plot them in SS limit in  Fig.~\ref{fig6:SSpotandenergyrelations}. The shaded area  indicates the intersection of the required conditions. The first condition, given with  Eq.~(\ref{SSbirincikosul}), puts the energy eigenvalues in a limited interval.  The other condition, given by  Eq.~(\ref{SSikincikosul}), assigns the minimum value of energy spectrum.
If a  KG particle with a rest mass energy $m_oc^2$ is confined in a GSWSP energy well that has the depth parameter less than $\frac{m_0c^2}{2}$, only positive bound states can be obtained. Negative energy eigenvalues start to appear with constriction, in GSWSP energy wells that have  $V_0$  values in between $m_0c^2$ and $\frac{m_0c^2}{2}$. In a GSWSP energy well that is constituted with $V_0=m_0c^2$ value, the whole range of energy spectrum from $-m_0c^2$ to $m_0c^2$ is obtained. For other potential energy wells that have greater values of $V_0$ beyond $m_0c^2$, the energy spectrum  interval does not enlarge further.

\subsubsection{Pseudo-spin symmetric limit}

\begin{figure}[!htb]
\centering
\includegraphics[totalheight=0.45\textheight]{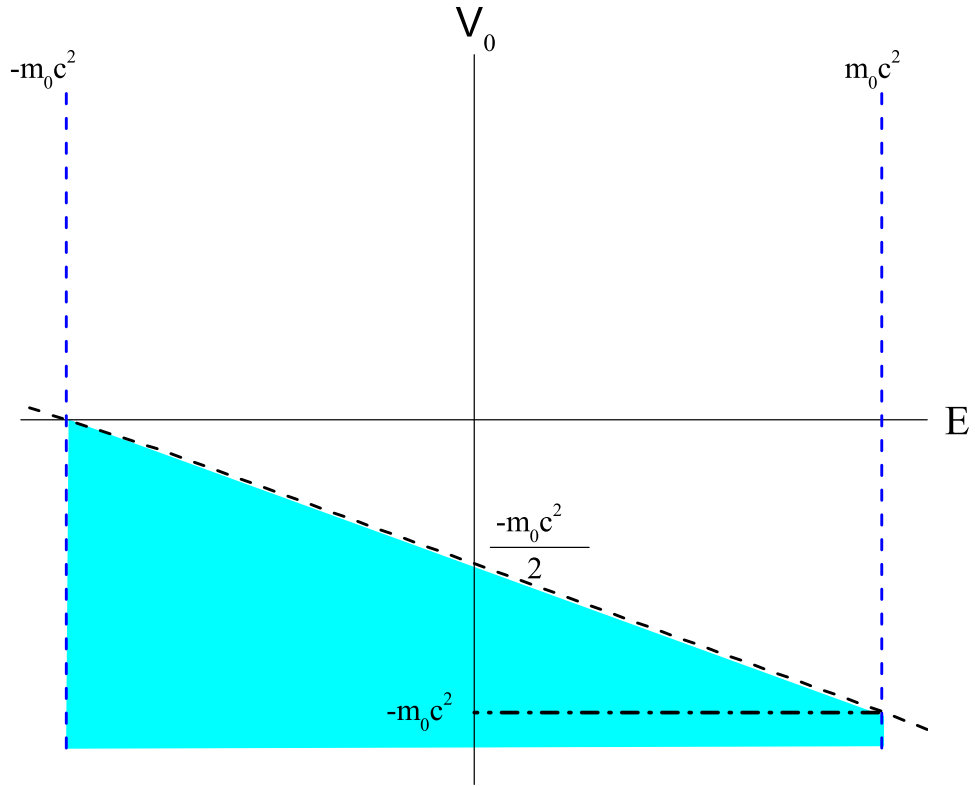}
   \caption{Possible energy eigenvalue region for a confined particle in PSS limit. Since potential depth parameter defined positively, despite from
   the SS limit, a particle cannot be confined in the PSS limit.} \label{fig7:PSSpotandenergyrelations}
\end{figure}

For the PSS limit, the analysis of the required conditions given in  the subsection~\ref{Bound State Conditions} ends up with a surprising result. The inequalities given with  Eq.~(\ref{SSbirincikosul}) and Eq.~(\ref{SSikincikosul}) are satisfied in the shaded area plotted in Fig.~\ref{fig7:PSSpotandenergyrelations}. On the other hand, a GSWSP energy well occurs only for a positive potential depth parameter $V_0$. Therefore,  the energy eigenvalue solution of bound state in the PSS limit is an empty set and consequently, a KG particle cannot be confined within a GSWSP well.

\subsection{Asymptotic Behaviors}\label{ASB}

In order to analyze the asymptotic behaviours of the wave
functions, the transformation variables $z$ and $y$ are examined
at positive and negative infinities and found to be zero. Thus,
the asymptotic behavior of hypergeometric functions gives unity.
The other multipliers, namely $(z-1)^{\nu_{\mp}}$ and
$(y-1)^{\nu_{\mp}}$, act like phase multipliers and yield to
$e^{-\frac{\pi \kappa_{\mp}}{\alpha}}$. Since  the wave number
$k_{\mp}$ are determined as a  positive real number by the imposed
conditions studied in  the Eq.~(\ref{SSbirincikosul}), the wave
functions behave  as exponential functions.
\begin{eqnarray}
 \phi_L^\mp(x\rightarrow -\infty) &\approx& \Big[D_1^\mp e^{x k_\mp } + D_2^\mp e^{-xk_\mp } \Big]e^{-\frac{\pi \kappa_{\mp}}{\alpha}} , \\
 \phi_R^\mp(x\rightarrow \infty) & \approx & \Big[D_3^\mp e^{-xk_\mp} + D_4^\mp  e^{xk_\mp } \Big]e^{-\frac{\pi \kappa_{\mp}}{\alpha}}.
\end{eqnarray}
To be consistent with being a solution to a bound state problem, $D_2^\mp$ and  $D_4^\mp$ should be taken as zero. Thus, the wave functions vanish
at both infinities.

Finally, it should be emphasized that the wave functions do not depend on the surface terms at infinities as one can expect.

\subsection{The Continuity Conditions}\label{CC}

The wave function should be continuous and well-defined at every point. Therefore, the matching of the solutions at the critical point $x=0$ should
be investigated.
\begin{eqnarray}
  \phi_L^\mp(x) \bigg|_{x=0^-} &=& \phi_R^\mp(x) \bigg|_{x=0^+},  \\
  \frac{d \phi_L^\mp(x)}{dx} \bigg|_{x=0^-} &=& \frac{d \phi_R^\mp(x)}{dx} \bigg|_{x=0^+}.
\end{eqnarray}
Note that at this critical point two transformations that have been used, namely,  $z\Big|_{x\rightarrow 0^-}$ and $y\Big|_{x\rightarrow 0^+}$, yield
 to a same non-zero number, hereby $t_0$,
\begin{eqnarray}
  t_0 \equiv (1+e^{-\alpha L})^{-1}.
\end{eqnarray}
From the equality of the wave function
\begin{eqnarray}
(D_1^{\mp}-D_3^\mp)t_0^{\mu_\mp}(t_0-1)^{\nu_\mp}M_1^{\mp}&=&0, \label{kendisi}
\end{eqnarray}
 is found and from the equality of the derivation of the wave function
\begin{eqnarray}
(D_1^{\mp}+D_3^\mp)t_0^{\mu_\mp}(t_0-1)^{\nu_\mp}\Bigg[\Bigg(\frac{\mu_\mp}{t_0}+\frac{\nu_\mp}{t_0-1}\Bigg)M_1^{\mp}+
\frac{(\mu_\mp+\theta_\mp+\nu_\mp)(1+\mu_\mp-\theta_\mp+\nu_\mp)}{1+2\mu_\mp} M_3^{\mp}\Bigg] =0, \nonumber \\ \,\,\,\,\,\,\,\,\, \label{turevi}
\end{eqnarray}
is obtained. Here
\begin{eqnarray}
  M_1^\pm &\equiv& \,\,\, {}_2F_1[\mu_\mp+\theta_\mp+\nu_\mp,1+\mu_\mp-\theta_\mp+\nu_\mp,1+2\mu_\mp;t_0], \\
  M_3^\pm &\equiv& \,\,\, {}_2F_1[1+\mu_\mp+\theta_\mp+\nu_\mp,2+\mu_\mp-\theta_\mp+\nu_\mp,2+2\mu_\mp;t_0].
\end{eqnarray}
Since $t_0$ is nearly equal to one, the hypergeometric functions have to be transformed. The identity given in  \cite{RefGradshytenRyzhikBook}
\begin{eqnarray}\label{HypergeoZ=1}
  _2F_1(a,b,c;t) &=& \frac{\Gamma(c)\Gamma(c-a-b)}{\Gamma(c-a)\Gamma(c-b)} \,\,\, _2F_1(a,b, a+b-c+1;1-t)+(1-t)^{c-a-b}\nonumber \\
  &&\times \frac{\Gamma(c)\Gamma(a+b-c)}{\Gamma(a)\Gamma(b)}\,\,\, _2F_1(c-a,c-b,
  c-a-b+1;1-t). \,\,\,\,\,\,\,\,\,\,\,\,
\end{eqnarray}
is used and derived
\begin{eqnarray}
  M_1^\mp &=& S_1^\mp N_1^\mp+ (1-t_0)^{-2\nu_\mp}S_2^\mp N_2^\mp, \\
  M_3^\mp &=& S_3^\mp N_3^\mp+ (1-t_0)^{-1-2\nu_\mp}S_4^\mp N_4^\mp.
\end{eqnarray}
with the new definitions
\begin{eqnarray}
  S_1^\mp &\equiv& \frac{\Gamma(1+2\mu_\mp)\Gamma(-2\nu_\mp)}{\Gamma(1+\mu_\mp-\theta_\mp-\nu_\mp)\Gamma(\mu_\mp+\theta_\mp-\nu_\mp)}, \\
  S_2^\mp &\equiv& \frac{\Gamma(1+2\mu_\mp)\Gamma(2\nu_\mp)}{\Gamma(1+\mu_\mp-\theta_\mp+\nu_\mp)\Gamma(\mu_\mp+\theta_\mp+\nu_\mp)}, \\
  S_3^\mp &\equiv& \frac{\Gamma(2+2\mu_\mp)\Gamma(-1-2\nu_\mp)}{\Gamma(1+\mu_\mp-\theta_\mp-\nu_\mp)\Gamma(\mu_\mp+\theta_\mp-\nu_\mp)}, \\
  S_4^\mp &\equiv& \frac{\Gamma(2+2\mu_\mp)\Gamma(1+2\nu_\mp)}{\Gamma(2+\mu_\mp-\theta_\mp+\nu_\mp)\Gamma(1+\mu_\mp+\theta_\mp+\nu_\mp)},
\end{eqnarray}
and
\begin{eqnarray}
  N_1^\mp &\equiv&  \,\,\, {}_2F_1[\mu_\mp+\theta_\mp+\nu_\mp,1+\mu_\mp-\theta_\mp+\nu_\mp,1+2\nu_\mp;1-t_0],  \\
  N_2^\mp &\equiv&  \,\,\, {}_2F_1[1+\mu_\mp-\theta_\mp-\nu_\mp,\mu_\mp+\theta_\mp-\nu_\mp,1-2\nu_\mp;1-t_0],  \\
  N_3^\mp &\equiv&  \,\,\, {}_2F_1[1+\mu_\mp+\theta_\mp+\nu_\mp,2+\mu_\mp-\theta_\mp+\nu_\mp,2+2\nu_\mp;1-t_0],  \\
  N_4^\mp &\equiv&  \,\,\, {}_2F_1[1+\mu_\mp-\theta_\mp-\nu_\mp,\mu_\mp+\theta_\mp-\nu_\mp,-2\nu_\mp;1-t_0].
\end{eqnarray}
Note that the behavior of the wave function solutions  within the
GSWSP well, to the contrary to the behavior at infinities, depends
on the surface effects by $\theta^\mp$ terms as suggested.
\subsection{Quantization and the Wave Function Solutions}\label{Quan}
The continuity conditions give the quantized energy spectra with the solutions of the Eq.~(\ref{kendisi}) and Eq.~(\ref{turevi}). As a consequence of
the combined solutions the energy spectra are separated into two subsets, namely even, $E_n^e$, and odd, $E_n^o$, subsets respectively.
\subsubsection{Even Solutions}
The even solutions are obtained by the equality of $D_1^\mp$ and $D_3^\mp$ with
\begin{eqnarray}
  \frac{(S_1^\mp N_1^\mp)+(1-t_0)^{-2\nu_\mp}(S_2^\mp N_2^\mp)}{(S_3^\mp N_3^\mp)+(1-t_0)^{-1-2\nu_\mp}(S_4^\mp N_4^\mp)}&=&  -\frac{(\mu_\mp+\theta_\mp+\nu_\mp)(1+\mu_\mp-\theta_\mp+\nu_\mp)t_0(t_0-1)}
  {(1+2\mu_\mp)\big((\mu_\mp+\nu_\mp)t_0-\mu_\mp\big)}.\,\,\, \label{ciftcozum}
\end{eqnarray}
The wave function
\begin{eqnarray}
  \phi_L^\mp(z) &=& D_1^\mp z^{\mu_\mp} (z-1)^{\nu_\mp} \,\,\, {}_2F_1[\mu_\mp+\theta_\mp+\nu_\mp,1+\mu_\mp-\theta_\mp+\nu_\mp,1+2\mu_\mp;z],\\
  \phi_R^\mp(y) &=& D_1^\mp y^{\mu_\mp} (y-1)^{\nu_\mp} \,\,\, {}_2F_1[\mu_\mp+\theta_\mp+\nu_\mp,1+\mu_\mp-\theta_\mp+\nu_\mp,1+2\mu_\mp;y],
\end{eqnarray}
guarantees a non-zero value at critical point
\begin{eqnarray}
  \phi_L^\mp(t_0) &=&  \phi_R^\mp(t_0)=D_1^\mp t_0^{\mu_\mp} (t_0-1)^{\nu_\mp} M_1^\mp.
\end{eqnarray}

\subsubsection{Odd Solutions}
The odd solutions are found by the combined solutions of $D_1^\mp+D_3^\mp=0$ and
\begin{eqnarray}
  \frac{S_1^\mp N_1^\mp}{S_2^\mp N_2^\mp} &=&
  -(1-t_0)^{-2\nu_\mp}.
\end{eqnarray}
The odd wave function
\begin{eqnarray}
  \phi_L^\mp(z) &=& D_1^\mp z^{\mu_\mp} (z-1)^{\nu_\mp} \,\,\, {}_2F_1[\mu_\mp+\theta_\mp+\nu_\mp,1+\mu_\mp-\theta_\mp+\nu_\mp,1+2\mu_\mp;z],\\
  \phi_R^\mp(y) &=& -D_1^\mp y^{\mu_\mp} (y-1)^{\nu_\mp} \,\,\, {}_2F_1[\mu_\mp+\theta_\mp+\nu_\mp,1+\mu_\mp-\theta_\mp+\nu_\mp,1+2\mu_\mp;y],
  \label{tekcozum}
\end{eqnarray}
is zero at the critical point
\begin{eqnarray}
  \phi_L^\mp(t_0) &=&   D_1^\mp t_0^{\mu_\mp} (t_0-1)^{\nu_\mp} M_1^\mp,\\
  \phi_R^\mp(t_0) &=&  -D_1^\mp t_0^{\mu_\mp} (t_0-1)^{\nu_\mp} M_1^\mp.
\end{eqnarray}
since $M_1^\mp=0$.

Remark that the term that indicates the surface interactions, namely $W$, does not take part in the wave numbers related with $\mu$ and $\nu$ parameters, although it is present in the coefficients $\beta$ and $\gamma$. On the other hand, $W$ stands in the hypergeometric functions. With this point of view, to have pocket or barrier in the potential energy function is not related with the wave numbers while it is related with the particle's energy, rest mass and the potential depth parameter.

\section{Applications} \label{sec:uygulama}
In the previous section, we  algebraically proved that bound states can occur only in the presence of the SS limit. In this section,  numerical
results are dealt with to strengthen the obtained conclusions for two different types of GSWSP energy wells.

\subsection{Bound states of GSWSP with repulsive surface effects}

In this subsection,   a confined neutral Kaon, where its rest mass energy is $m_0c^2=497.648 \,\,MeV$, is initially considered in  a GSWSP energy well constructed with  arbitrarily determined parameters   $V_0=\frac{m_0c^2}{2}$, $W=2 m_0c^2$, $\alpha=1\,\, fm^{-1}$ and $L=6\,\, fm$. Note that, $\alpha L=6$ which makes $e^{\alpha L}$  much  greater than one. The positive value of $W$ indicates that the considered surface forces are repulsive. Moreover, a surface barrier occurs since $V_0 < W$ and its height is found to be $\frac{9}{32} m_0c^2$ ($139.963 \,\, MeV$).

The Newton-Raphson (NR) method is used to calculate the energy spectra via the  transcendental Eq.~(\ref{ciftcozum}) and Eq.~(\ref{tekcozum}). The obtained eigenvalues are tabulated in Table~\ref{tab1}. with their corresponding node numbers denoted by $n$.

\begin {table}[!ht]
\caption{\label{tab1}
Energy spectrum of a confined neutral Kaon in a GSWSP energy well with repulsive surface interactions. Note that all calculated energies have units in $MeV$. }
\begin{tabular}{|c|c|c|c|c|c|c|c|c|c|}
  \hline
  $n$  & 0& 1 & 2  & 3 & 4 & 5 & 6 &  7   & 8    \\ \hline
  $E_n$ & 33.962  & 86.193  & 139.950  & 194.935   & 249.319  & 303.138 & 355.809 & 407.437 & 457.746 \\
    \hline
\end{tabular}
%\label{ss table 1}
% \caption{The units of $\varepsilon^+$ and energy spectra are $MeV$.}
\end{table}

Using the first three eigenvalues, their corresponding unnormalized eigenfunctions have been plotted in Fig.~\ref{fig8:barierwavefunc}. The repulsive surface effects, in addition to the bulk effect, push the particle toward the center. Therefore, the probability density near the core is higher.

The procedure is repeated to calculate bound state energy spectra for different GSWSP energy wells. All parameters are kept same except the potential depth parameter. The spectra are tabulated in Table~\ref{tab2}  for the cases $V_0=m_0c^2$, $V_0=\frac{3m_0c^2}{2}$, respectively.

\begin {table}[!ht]
\caption{\label{tab2}
Energy spectra of a confined neutral Kaon in  GSWSP energy wells that have been parameterized with $W=2 m_0c^2$, $\alpha=1\,\, fm^{-1}$ and $L=6\,\, fm$. Note that all calculated energies have units in $MeV$ and $n$ are the node numbers. }
\begin{tabular}{|l|c|c|l|c|c|}
  \hline
  n &    $V_0= m_0c^2$    & $V_0= \frac{3m_0c^2 }{2}$    & n    & $V_0= m_0c^2$    & $V_0= \frac{3m_0c^2 }{2}$        \\\hline
  0                     & -410.270       & -493.730     & 11    & 297.283 & 59.956    \\
  1                     & -320.136       & -467.612     & 12    & 347.212 & 110.177   \\
  2                     & -242.476       & -424.559     & 13    & 395.518 & 159.275   \\
  3                     & -170.674       & -373.166     & 14    & 442.678 & 207.153   \\
  4                     & -103.544       & -318.552     & 15    & 486.799 & 253.807   \\
  5                     & -39.538        & -262.729     & 16    &         & 299.119   \\
  6                     & 21.623         & -207.001     & 17    &         & 343.019   \\
  7                     & 80.580         & -151.731     & 18    &         & 385.311   \\
  8                     & 137.411        & -97.334      & 19    &         & 425.766   \\
  9                     & 192.442        & -43.837      & 20    &         & 463.891   \\
  10                    & 245.676        & 8.581        & 21    &         &           \\
  \hline
\end{tabular}
%\label{ss table 1}
% \caption{The units of $\varepsilon^+$ and energy spectra are $MeV$.}
\end{table}

The calculated energy spectra are plotted versus the depth parameters of the potential energy well and node number in Fig.~\ref{fig9:KGboundEnergies}. Whether a neutral Kaon is confined in the GSWSP energy well is determined by the potential parameter  value $\frac{m_0c^2}{2}$. Only nine eigenvalues are obtained. Note that, they have only positive values as predicted in Fig.~\ref{fig6:SSpotandenergyrelations}. In a deeper GSWSP energy well, $V_0=m_0c^2$, the number of the possible microstate numbers increases. Moreover, the range of the energy spectrum widens to be in between $-m_0c^2$ to $m_0c^2$, as expected. Sixteen node numbers increase to twenty one when the GSWSP energy well becomes deeper by setting the depth parameter as $\frac{3m_0c^2}{2}$. In this case, the repulsive surface effects become less dominant on the bulk effects. The height of the barrier is calculated to be $15.552 \,\, MeV$. Note that, the energy eigenvalues are only found to be in the KG interval.

\subsection{Bound states of GSWSP with attractive surface effects}
In this subsection, a neutral Kaon confinement is considered in a GSWSP energy well constructed with the attractive surface effects, hence $W$ is  negative and equal to $-2 m_0c^2$. The other parameters namely $V_0$, $\alpha$ and $L$ are chosen to be same as those  in the previous subsection and equal to $\frac{m_0c^2}{2}$, $1\,\, fm^{-1}$ and $6\,\, fm$, respectively. The depth of the local pockets near the surface is calculated to be $-388.788 \,\,
 MeV$.

The energy spectrum, that is calculated with the NR method  is tabulated in Table~\ref{tab3} with their corresponding node numbers. The bound state energy spectrum consists of sixteen eigenvalues. In the spectrum the initial values of node number is four rather than zero, which means the confined particle  is not in the ground state. A particle in its ground level is expected to be localized around the center of the potential well, whereas, neutral Kaon does not satisfy this criterion, since the surface effect deepens the well around the surface. The particle is localized at a sufficiently excited level with a non-zero $n$ value.  Therefore, the particle is expected to localize not only around the center of the well. The probability density distribution extends beyond the nuclear surface as demonstrated in Fig.~\ref{fig10:ProbabilityPocketEpsilon=0_50-n=678_456}, implying a non-zero probability in outer vicinity of the well as a quantum mechanical effect.

\begin {table}[!ht]
\caption{\label{tab3}
Energy spectrum of a confined neutral Kaon in a GSWSP energy well with attractive surface forces. The potential well has parameters $V_0=\frac{m_0c^2}{2}$, $W=-2 m_0c^2$, $\alpha=1\,\, fm^{-1}$ and $L=6\,\, fm$. Note that all calculated energies have units in $MeV$.}
\begin{tabular}{|c|c|c|c|c|c|c|c|}
  \hline
  n    &$E_n$     &n & $E_n$      &  n   & $E_n$   & n      & $E_n$     \\\hline
  6    & 47.403   &10& 177.494    &  14  & 305.153 &  18    & 417.433   \\
  7    & 78.700   &11& 210.232    &  15  & 335.082 &  19    & 441.465   \\
  8    & 111.445  &12& 242.588    &  16  & 371.316 &  20    & 463.049   \\
  9    & 144.380  &13& 274.238    &  17  & 391.456 &  21    & 481.232   \\
  \hline
\end{tabular}
%\label{ss table 1}
% \caption{The units of $\varepsilon^+$ and energy spectra are $MeV$.}
\end{table}

\subsection{Non-existence of bound state solutions numerically in the PSS limit}

The parameters defined for the attractive and the repulsive surface effects in the previous subsections are used in PSS limit to investigate a
bound
state solution numerically. The NR method verify that there is no bound state energy spectrum in PSS limit.

\section{Conclusion}\label{sec:sonuc}

In this work, in the limits of  spin symmetry (SS) and pseudospin symmmetry (PSS), the bound state solutions of the Klein-Gordon(KG) equation are investigated in the generalized symmetric Woods-Saxon potential (GSWSP) energy in one spatial dimension. The GSWSP differs from the usual Woods-Saxon potential by the additional repulsive or the attractive surface interactions terms. In the first part of the manuscript the structure of the GSWSP energy is investigated in detail. Then, the KG equation is obtained briefly in the SS and PSS symmetry limits. The conditions
on the wave numbers are derived and it is shown that an energy spectrum can be obtained only in the SS limit. To have numerical results, in SS limit, a neutral Kaon is chosen to be confined in repulsive and attractive GSWSP energy. The energy spectra are calculated for both cases. Moreover, in the PSS limit, it is verified  also numerically that an energy spectrum does not exist.

\section*{Acknowledgments}
This work was partially supported by the Turkish Science and
Research Council (T\"{U}B\.{I}TAK) and Akdeniz University. The author thanks to Dr. M. Erdo\u{g}an for the discussion on the preparation of the manuscript and to Dr. E. Pehlivan for the  proof reading of the first version and  to Prof. M. Horta\c{c}su  for the  proof reading of the final version. Furthermore, the author would like to emphasize his gratitude to the anonymous referee for her/his careful review of the manuscript with very kind comments and suggestions.

%\newpage

\section{References}
%%%%%%%%%%%%%%%%%%%%%%%%%%%%%%%%%%%%%%%%%%%%%%%%%%%%%%%%%%%%%%%%%%%%%%%%%%%%%%%%%%%%%%%%%%%%%%%%%%%%%%%%%%%%%
%%%%%%%%%%%%%%%%%%%%%%%%%%%%%%%%%%%%%%%%%%%%%%%%%%%%%%%%%%%%%%%%%%%%%%%%%%%%%%%%%%%%%%%%%%%%%%%%%%%%%%%%%%%%%

\newpage

\newpage

\newpage

\newpage

\newpage

\newpage

\newpage

\newpage
\begin{figure}[!htb]
\centering
\includegraphics[totalheight=0.45\textheight]{fig8.eps}
   \caption{The first three unnormalized eigenfunctions in the presence of repulsive surface effects in the SS limits.}
   \label{fig8:barierwavefunc}
\end{figure}

\newpage
\begin{figure}[!htb]
\centering
\includegraphics[totalheight=0.40\textheight]{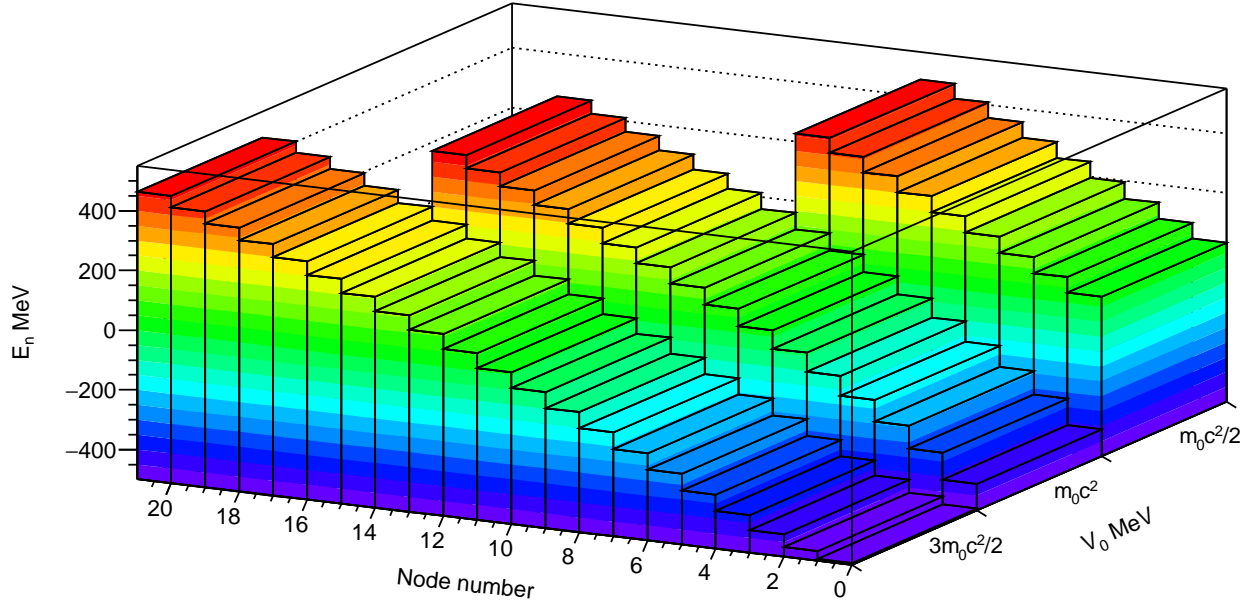}
   \caption{Bound state's energy spectra for different GSWSP energy wells versus the node number in the SS limit. Note that $0.5 m_oc^2=248.824$ $MeV$.} \label{fig9:KGboundEnergies}
\end{figure}

\newpage
\begin{figure}[!htb]
\centering
\includegraphics[totalheight=0.45\textheight]{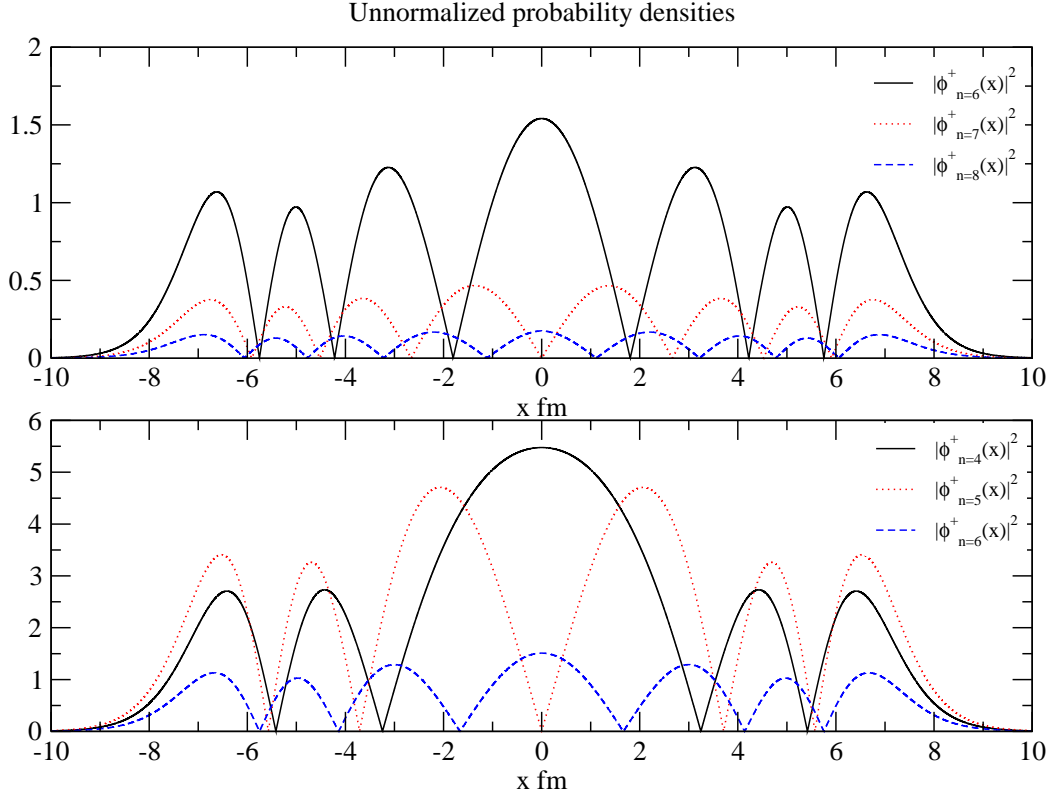}
   \caption{Unnormalized probability densities of the confined neutral Kaon in the GSWSP well under attractive surface   effects. Note that there is a non zero probability that the Kaon can be confined outside the nucleus  near the surface which is classically    forbidden.} \label{fig10:ProbabilityPocketEpsilon=0_50-n=678_456}
\end{figure}

\end{document}